# LABEL TRANSFER FROM APOGEE TO LAMOST: PRECISE STELLAR PARAMETERS FOR 450,000 LAMOST GIANTS

Anna Y. Q. Ho,[1,2] Melissa K. Ness,[2] David W. Hogg,[2,3,4,5]
Hans-Walter Rix,[2] Chao Liu,[6] Fan Yang,[6] Yong Zhang,[7] Yonghui Hou,[7] and Yuefei Wang[7]

[1]*Cahill Center for Astrophysics, California Institute of Technology, MC 249-17, 1200 E California Blvd, Pasadena, CA, 91125, USA*

[2]*Max-Planck-Institut für Astronomie, Königstuhl 17, D-69117 Heidelberg, Germany*

[3]*Simons Center for Data Analysis, 160 Fifth Avenue, 7th floor, New York, NY 10010, USA*

[4]*Center for Cosmology and Particle Physics, Department of Phyics, New York University, 4 Washington Pl., room 424, New York, NY, 10003, USA*

[5]*Center for Data Science, New York University, 726 Broadway, 7th floor, New York, NY 10003, USA*

[6]*Key Laboratory of Optical Astronomy, National Astronomical Observatories, Chinese Academy of Sciences, Datun Road 20A, Beijing 100012, China*

[7]*Nanjing Institute of Astronomical Optics & Technology, National Astronomical Observatories, Chinese Academy of Sciences, Nanjing 210042, China*

## ABSTRACT

In this era of large-scale stellar spectroscopic surveys, measurements of stellar attributes ("labels," i.e. parameters and abundances) must be made precise and consistent across surveys. Here, we demonstrate that this can be achieved by a data-driven approach to spectral modeling. With *The Cannon*, we transfer information from the APOGEE survey to determine precise $T_{\rm eff}$, $\log g$, [Fe/H], and [$\alpha$/M] from the spectra of 450,000 LAMOST giants. *The Cannon* fits a predictive model for LAMOST spectra using 9952 stars observed in common between the two surveys, taking five labels from APOGEE DR12 as ground truth: $T_{\rm eff}$, $\log g$, [Fe/H], [$\alpha$/M], and K-band extinction $A_{\rm k}$. The model is then used to infer $T_{\rm eff}$, $\log g$, [Fe/H], and [$\alpha$/M] for 454,180 giants, 20% of the LAMOST DR2 stellar sample. These are the first [$\alpha$/M] values for the full set of LAMOST giants, and the largest catalog of [$\alpha$/M] for giant stars to date. Furthermore, these labels are by construction on the APOGEE label scale; for spectra with S/N > 50, cross-validation of the model yields typical uncertainties of 70 K in $T_{\rm eff}$, 0.1 in $\log g$, 0.1 in [Fe/H], and 0.04 in [$\alpha$/M], values comparable to the broadly stated, conservative APOGEE DR12 uncertainties. Thus, by using "label transfer"

ah@astro.caltech.edu

to tie low-resolution (LAMOST R $\sim$ 1800) spectra to the label scale of a much higher-resolution (APOGEE R $\sim$ 22,500) survey, we substantially reduce the inconsistencies between labels measured by the individual survey pipelines. This demonstrates that label transfer with *The Cannon* can successfully bring different surveys onto the same physical scale.

## 1. LABEL TRANSFER USING *The Cannon*

A diverse suite of large-scale spectroscopic stellar surveys have been measuring spectra for hundreds of thousands of stars in the Milky Way. Among them are APOGEE (Majewski et al. 2015), Gaia-ESO (Gilmore et al. 2012), GALAH (De Silva et al. 2015), LAMOST (Zhao et al. 2012), RAVE (Kordopatis et al. 2013), SEGUE (Yanny et al. 2009), and *Gaia* (Gaia Collaboration 2016) with its radial velocity spectrometer. Stellar spectra are also obtained by surveys as side products: for example, SDSS has many more stellar spectra beyond SEGUE, obtained in the original survey and subsequent (non-SEGUE) phases like BOSS and eBOSS.

These surveys target different types of stars, in different parts of the sky, and at different wavelengths. For example, APOGEE observes in the near-infrared (near-IR) and targets predominantly giants in the dust-obscured mid-plane of the Galaxy, whereas GALAH observes in the optical and targets predominantly nearby main sequence stars. In addition, they observe at different resolutions and employ different data analysis methodologies for using spectra to derive a set of labels characterizing each star. In our work, we use the term "label" to collectively describe the full set of stellar attributes, e.g. physical parameters and element abundances like $T_{eff}$, $\log g$, $[\alpha/M]$, and $[X/H]$. We adopt this term from the supervised machine learning literature, as our methodology (*The Cannon*) is an adaptation of supervised learning to suit the particulars of stellar spectra.

The suite of spectroscopic surveys are complementary in their spatial coverage and scientific motivation, and there is enormous scientific promise in combining their results. However, diversity is also the reason why surveys cannot be rigorously stitched together at present: different pipelines measure substantially different labels for the same stars (e.g. Smiljanic et al. (2014)). For example, Chen et al. (2015) compared the three stellar parameters $T_{eff}$, $\log g$, and $[Fe/H]$ between APOGEE and LAMOST, two of the most ambitious ongoing surveys, and found consistency in the photometrically-calibrated $T_{eff}$ but systematic biases in $\log g$ and $[Fe/H]$, as Figure 1 shows for 9952 objects observed and analyzed by both surveys. Furthermore, when Lee et al. (2015) used the SEGUE pipeline to measure parameters (including $[\alpha/M]$ and $[C/Fe]$) from LAMOST spectra, they found that the physical scale of SEGUE labels is systematically offset from that of other surveys, like APOGEE. The SEGUE pipeline could only be straightforwardly applied to the LAMOST spectra because the two surveys are qualitatively very similar, e.g. in their resolution and wavelength coverage.

Although such systematic label offsets may not be surprising for two surveys with disjoint wavelength coverage and very different spectral resolutions (see Section 2), labels are ultimately characteristics of stars and not of observations, and must therefore be unbiased and consistent between surveys to within the stated error bars. To that end, better techniques must be developed for bringing different surveys onto the same label scale.

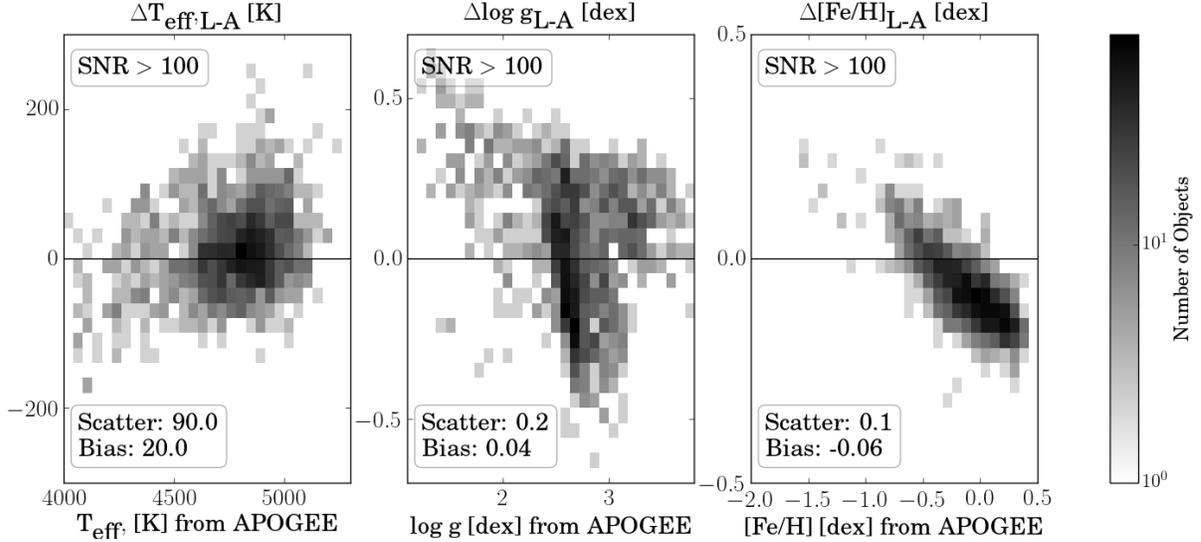

**Figure 1.** Systematic offsets in the labels $T_{eff}$, $\log g$, and [Fe/H] that were derived by the LAMOST ("L") and APOGEE ("A") pipelines, respectively. There are significant biases in $\log g$ and [Fe/H]. Shown for the 2183 stars that have been observed and analyzed by both surveys, and that have LAMOST spectra with S/N > 100. S/N values were calculated for each spectrum by taking the median of the flux-uncertainty ratio across all pixels.

We approach this problem of inter-survey systematic biases by using *The Cannon* (Ness et al. 2015), a new data-driven method for measuring stellar labels from stellar spectra in the context of large spectroscopic surveys. Ness et al. (2015) describe the method in detail; we direct the reader to this paper for details on what distinguishes this particular data-driven technique from others, and more specifically what distinguishes it from the MATISSE method (Recio-Blanco et al. 2006). Here, we recapitulate the fundamental assumptions and steps of *The Cannon* in the context of bringing surveys onto the same scale, and describe the procedure more concretely in sections 3 and 4.

Presume that Survey X and Survey Y are two spectral surveys that are not (yet) on the same label scale: their individual pipelines measure inconsistent labels for objects observed in common, as in Figure 1. Presume further that there are good reasons to trust the labels of Survey X more than those of Survey Y. This could be, for example, because Survey X has higher spectral resolution and higher S/N. Our goal is to resolve the systematic inconsistencies by bringing Survey Y onto Survey X's label scale. Ultimately, we want a model that can directly infer labels from Survey Y's spectra that are consistent with what *would* be measured by the Survey X pipeline from the corresponding Survey X spectra.

*The Cannon* relies on a few key assumptions: that stars with identical labels have very similar spectra, and that spectra vary smoothly with label changes. In other words, the continuum-normalized flux at each pixel in a spectrum is a smooth function of the labels that describe the object. The function that takes the labels and predicts the flux at each wavelength of the spectrum is called the "spectral model"; fitting for the coefficients of the spectral model is the goal of the first step, the "training" step.

In the training step, *The Cannon* uses the objects with spectra from Survey Y, and corresponding "reference labels" from Survey X for those objects, to fit for the spectral model coefficients at each pixel of the spectrum independently. The spectral model characterizes the flux at each pixel of a Survey Y spectrum as a function of corresponding Survey X labels, and predicts what the spectrum of an object observed in Survey Y would look like given a set of labels from Survey X.

In the second step, the "test" step, this model is used to derive likely labels for any (similar) object given its spectrum from Survey Y, including those not observed by Survey X. Note that if the Survey X pipeline has measured a dozen labels precisely and the Survey Y pipeline has only measured three, we can in principle use our model to infer extra, previously unknown labels from Survey Y spectra; we dub this process of transferring knowledge of labels from one survey to another "label transfer." Note also that in this approach, Survey X enters only through its labels, not the data (spectra, light curves, or otherwise) from which these labels were derived, and Survey Y enters only through its spectra. This distinguishes our approach from traditional cross-calibration techniques such as multi-linear fitting. Although the *outcome* of this process (consistent labels for a set of stars observed in common between two surveys) is the same as in traditional cross calibration, we make no use of the labels from the Survey Y pipeline. In a sense, cross-calibration is a byproduct of our label transfer analysis.

Note that this procedure does not require that the two surveys have any overlapping wavelength regions; indeed, that is one of its strengths. However, this also means that caution must be taken when transferring labels from one survey to another. One could imagine trying to measure a new label from a wavelength regime that has no sensitivity to that label. In that case, *The Cannon* could still "learn" to predict the label via astrophysical correlations with other labels. Thus, the model should always be inspected for astrophysical plausibility. The interpretability of the model is another strength of our approach, as addressed in Section 3 and especially Figure 5.

In this work, we take APOGEE to be Survey X and LAMOST to be Survey Y. We select APOGEE as the source of the trusted stellar labels because it is the higher-resolution survey ($R \approx 22,500$ versus $R \approx 1,800$ for LAMOST). We use four post-calibrated labels from APOGEE DR12, as measured by the ASPCAP pipeline (García Pérez et al. 2015): $T_{\text{eff}}$, $\log g$, [Fe/H], and [$\alpha$/M]. We also use the K-band extinction $A_k$; while not strictly an intrinsic property of the stars, it is a "label" in the sense that it is an immutable property of the stellar spectrum when observed from our location in the Galaxy. We decided to include extinction in constructing the model because the objects in the reference set (in the Galactic mid-plane) include visual extinctions up to $A_v \approx 3.5$ ($A_k \approx 0.4$). This impacts some of the optical spectra in the training step and in the test step, not only by reddening, but also by dust and gas absorption features.

Note that what we call [Fe/H] in this work is stored under the header PARAM_M_H in DR12. We use this value so that all four labels have gone through the same post-calibration procedure, but refer to it as [Fe/H] rather than [M/H] because it has been calibrated to the [Fe/H] of star clusters (Mészáros et al. 2013), and in order to be consistent with the terminology from LAMOST.

Of course, our key assumption - that stars with identical labels have very similar spectra - is only an approximation. In this case, we assume that any two stars with near-identical $T_{\text{eff}}$, $\log g$, [Fe/H], [$\alpha$/M], $A_k$ have near-identical spectra, regardless of spatial position (e.g. RA & Dec) or other properties (e.g. individual element abundances). This approximation should be a very good one, however, because the shape of each spectrum should be dominated by these five labels. This is supported by the quality of the model fit, e.g. as illustrated in Figure 10.

The 11,057 objects measured in common between APOGEE and LAMOST constitute the possible reference set for the training step; in practice, we use 9952 of these objects to fit for the spectral model, then apply this model to infer both new labels for the reference set, as well as labels for the remaining 444,228 LAMOST giants in DR2 *not* observed by APOGEE. By construction, these labels are tied to the APOGEE scale.

Like cross calibration techniques, our label transfer approach with *The Cannon* is fundamentally limited by the quality and breadth of the available reference set. In this case, the set of common objects happens to be entirely giants and we are therefore limited to applying our model to the giants in LAMOST DR2, which is why we must discard such a large fraction (80%) of our sample. Indeed, *The Cannon* model is only applicable within the label range in which it has been trained, and even then there is inevitably some extrapolation because we are not training on a set of labels that comprehensively describe a stellar spectrum. We return to this issue in Section 4, and direct the reader to Section 5.4 of Ness et al. (2015) for additional discussion of the issue of extrapolation in *The Cannon* and to Section 6 of Ness et al. (2015) for avenues for future improvement.

This work is an implementation of the general procedure that is described in detail in Ness et al. (2015). The primary distinguishing feature is how the LAMOST spectra were prepared for *The Cannon*, and we describe that process in Section 2.1. The fact that it performs well for spectra at very different wavelength regimes and resolutions illustrates the general applicability of this procedure to large uniform sets of stellar spectra, given a suitable reference set.

## 2. DATA: LAMOST SPECTRA AND APOGEE LABELS

The Large sky Area Multi-Object Spectroscopic Telescope (LAMOST) is a low-resolution ($R \approx 1,800$) optical ($3650 - 9000\,\text{Å}$) spectroscopic survey. The second data release (DR2; Luo A.L., Bai Z.R. et al. (2015)) is public and consists of spectra for over 4.1 million objects, as well as three stellar labels ($T_{\rm eff}$, $\log g$, [Fe/H]) for $\sim 2.2$ million stars. Although the survey does not select for a particular stellar type, many of the stars are red giants; the population of K giants numbers 500,000 in DR2 (Liu et al. 2014). Moreover, $>100,000$ red clump candidates have been identified in the DR2 catalog (Wan et al. 2015). Stellar labels for the LAMOST spectra are derived by the LAMOST Stellar Parameter pipeline (LASP; Wu et al. 2011a,b; Luo A.L., Bai Z.R. et al. 2015). LASP proceeds via two steps. In the first step, the Correlation Function Initial (CFI; Du et al. 2012) calculates the correlation coefficients between the measured spectrum and spectra from a synthetic grid, and finds the best match. This first-pass coarse estimate serves as the starting guess for the second step, which makes use of the Université de Lyon Spectroscopic Analysis Software (*ULySS*; Koleva et al. 2009; Wu et al. 2011b). In *ULySS*, each spectrum is fit to a grid of model spectra from the ELODIE spectral library (Prugniel & Soubiran 2001; Prugniel et al. 2007). These model spectra are a linear combination of non-linear components, optically convolved with a line-of-sight velocity distribution and multiplied by a polynomial function. Improved surface gravity values have been obtained for the metal-rich giant stars via cross-calibration with asteroseismically-derived values from *Kepler* (Liu et al. 2015).

APOGEE is a high-resolution ($R \approx 22,500$), high-S/N (S/N $\approx 100$), H-band (15200-16900 Å) spectroscopic survey, part of the Sloan Digital Sky Survey III (Majewski et al. (2015); Eisenstein et al. (2011)). Observations are conducted using a 300-fiber spectrograph (Wilson et al. 2010) on the 2.5 m Sloan Telescope (Gunn et al. 2006) at the Apache Point Observatory (APO) in Sunspot, New Mexico (USA) and consist primarily of red giants in the Milky Way bulge, disk, and halo. The most recent data release, DR12 (Alam et al. 2015; Holtzman et al. 2015), comprises spectra for $>100,000$ red giant stars together with their basic stellar parameters and 15 chemical abundances. The parameters and abundances are derived by the ASPCAP pipeline, which is based on chi-squared fitting of the data to 1D LTE models for seven labels: $T_{\rm eff}$, $\log g$, [Fe/H], [$\alpha$/M], [C/M], [N/M], and micro-turbulence (García Pérez et al. 2015). The best-matching synthetic spectrum for each star is found using the FERRE code (Allende Prieto et al. 2006).

## 2.1. *Preparing LAMOST Spectra for The Cannon*

To be used by *The Cannon*, any spectroscopic data set must satisfy the conditions laid out in Ness et al. (2015). The spectra must share a common line-spread function, be shifted to the rest-frame and sampled onto a common wavelength grid with uniform start and end wavelengths. The flux at each pixel of each spectrum must be accompanied by a flux variance that takes error sources such as photon noise and poor sky subtraction into account; bad data (e.g. regions with skylines and telluric regions) must be assigned inverse variances of zero or very close to zero. Finally, the spectra do not need to be continuum normalized, but they must be normalized in a consistent way that is independent of S/N; more precisely, the normalization procedure should be a linear operation on the data, so that it is unbiased as (symmetric) noise grows.

Preparatory steps were necessary to make the raw LAMOST spectra satisfy these criteria. First, the displacement from the rest-frame was calculated for each spectrum using the redshift value provided in the data file header, and the spectra shifted accordingly. (The redshift values are derived within the LAMOST data pipeline, from their cross-correlation procedure.) Spectra were then re-sampled onto the original grid using linear interpolation. After shifting, we applied lower and upper wavelength cuts and sampled all spectra onto a common wavelength grid spanning 3905 Å – 9000 Å. All of these operations were performed on both the flux and inverse variance arrays.

Each spectrum was normalized by dividing the flux at each $\lambda_0$ by $\bar{f}(\lambda_0)$, which was derived by an error-weighted, broad Gaussian smoothing:

$$\bar{f}(\lambda_0) = \frac{\sum_i (f_i \, \sigma_i^{-2} \, w_i(\lambda_0))}{\sum_i (\sigma_i^{-2} \, w_i(\lambda_0))}, \qquad (1)$$

where $f_i$ is the flux at pixel $i$, $\sigma_i$ is the uncertainty at pixel $i$, and the weight $w_i(\lambda_0)$ is drawn from a Gaussian

$$w_i(\lambda_0) = e^{-\frac{(\lambda_0 - \lambda_i)^2}{L^2}} \qquad (2)$$

$L$ was chosen to be 50 Å, much broader than typical atomic lines.

To emphasize, this "normalization" is in no sense "continuum normalization," and is different from the standard normalization used in spectral analysis. Our goal in preparing the spectra in this way is to simplify the modeling procedure by removing overall flux, flux calibration, and large-scale shape changes from the spectra.

The procedure is illustrated in Figure 2, which shows three spectra corresponding to a sample reference object: its APOGEE spectrum, its LAMOST spectrum overlaid with its Gaussian-smoothed "continuum," and final "normalized" LAMOST spectrum.

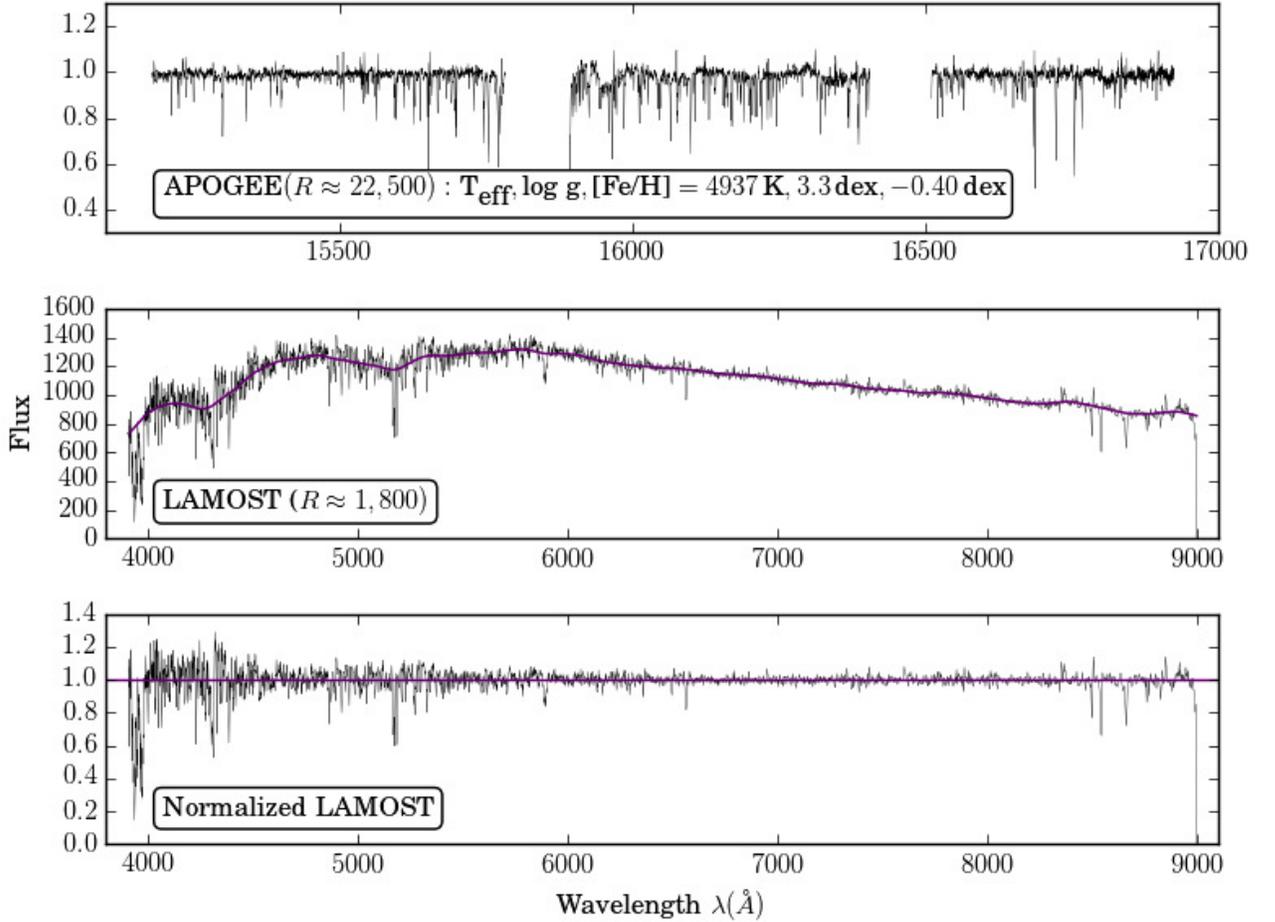

**Figure 2.** Spectra of a sample reference object (2MASS ID 2M07101078+2931576). The top panel shows the normalized APOGEE spectrum (with its basic stellar labels) and the middle panel shows the raw LAMOST spectrum overlaid with the Gaussian-smoothed version of itself. The bottom panel shows the resulting "normalized" spectrum, determined by dividing the black line by the purple line in the middle panel. *The Cannon* operates on the normalized spectrum in the bottom panel, although note that this "normalization" is different from the standard normalization used in spectral analysis. APOGEE and LAMOST spectra are qualitatively very different, in wavelength coverage and resolution.

## 3. *The Cannon* TRAINING STEP: MODELING LAMOST SPECTRA AS A FUNCTION OF APOGEE LABELS

In the training step, as described in Section 1, *The Cannon* uses objects observed in common between the two surveys of interest. These common objects, used to train the model, are called *reference objects.* For each reference object, *The Cannon* uses the spectra from one survey (in this case, LAMOST) and the corresponding "trusted labels" from the other survey (in this case, APOGEE). This data – spectra from one survey, labels from the other – is used to fit a predictive model independently at each wavelength of a (LAMOST) spectrum. Given a set of APOGEE labels, this model seeks to predict every pixel of a LAMOST spectrum for a star with those properties.

10To select reliable reference objects, we make a number of quality cuts to the full set of 11,057 objects in common between LAMOST DR2 and APOGEE DR12. We eliminate stars with unreliable $T_{\text{eff}}$, $\log g$, [Fe/H], [$\alpha$/M], or $A_k$ as described in Holtzman et al. (2015). This involves excising the 677 objects with $T_{\text{eff}} < 3500$ or $T_{\text{eff}} > 6000$, with [$\alpha$/M] $< 0.1$ dex, or with `ASPCAPFLAG` set. This leaves 10,380 objects.

Furthermore, a reliable reference object is by definition one that can be captured by the spectral model. So, we run an iteration of *The Cannon* on the 10,380 objects from the first cut: we train the model and use it to infer new labels for all 10,380 objects. We excise the 428 objects ($< 0.5\%$) whose difference from the reference (APOGEE) value in any label is greater than four times the scatter in that label. This leaves 9952 objects of the original 11,057. These cuts, sensible but somewhat *ad hoc*, still leave a very extensive set of reference objects.

The label space of the remaining reference set is well-sampled, as seen in Figures 3 and 4. Figure 3 shows the distribution of the remaining 9952 reference objects in (LAMOST $T_{\text{eff}}$, LAMOST $\log g$) label space. The black points in the background are the full LAMOST DR2 sample, with their values from the LAMOST pipeline. The overlaid colored points are the reference objects; in the left panel, they are shown with their LAMOST pipeline values, and in the right panel, they are shown with their APOGEE pipeline values. It is only the APOGEE labels, shown as colored dots in the right panel, that are used in the training step.

Figure 4 again shows the distribution of labels for the 9952 reference objects, this time for each label individually. The values from the LAMOST pipeline are shown in yellow and the corresponding values from the APOGEE pipeline are shown in purple. The APOGEE (purple) values comprise the reference set (are used to train the spectral model).

*The Cannon* uses the reference objects to fit for a spectral model that characterizes the flux in each pixel of the (normalized) spectrum as a function $g$ of the labels of the star. In this case, the flux $f_{n\lambda}^L$ for a spectrum $n$ at wavelength $\lambda$ in the LAMOST survey (L) can be written as

$$f_{n\lambda}^L = g(\boldsymbol{\ell}_n^A | \boldsymbol{\theta}_\lambda) + \text{noise} \tag{3}$$

where $\boldsymbol{\theta}_\lambda$ is the set of spectral model coefficients at each wavelength $\lambda$ of the LAMOST spectrum and $\boldsymbol{\ell}_n^A$ is some (possibly complex) function of the full set of labels from APOGEE. The noise model is noise $= [s_\lambda^2 + \sigma_{n\lambda}^2] \xi_{n\lambda}$, where each $\xi_{n\lambda}$ is a Gaussian random number with zero mean and unit variance. The noise is thus a root-mean-square (rms) combination of two contributions: the inherent uncertainty in the spectrum from e.g. instrument effects and finite photon counts ($\sigma_{n\lambda}$), and intrinsic scatter in the model at each wavelength ($s_\lambda$). This intrinsic scatter can be thought of as the expected deviation of the spectrum from the model at that pixel, even in the limit of vanishing measurement uncertainty. Handling uncertainties by fitting for a noise model independently at each pixel is a key feature of *The Cannon* and distinguishes it from traditional machine learning methods.

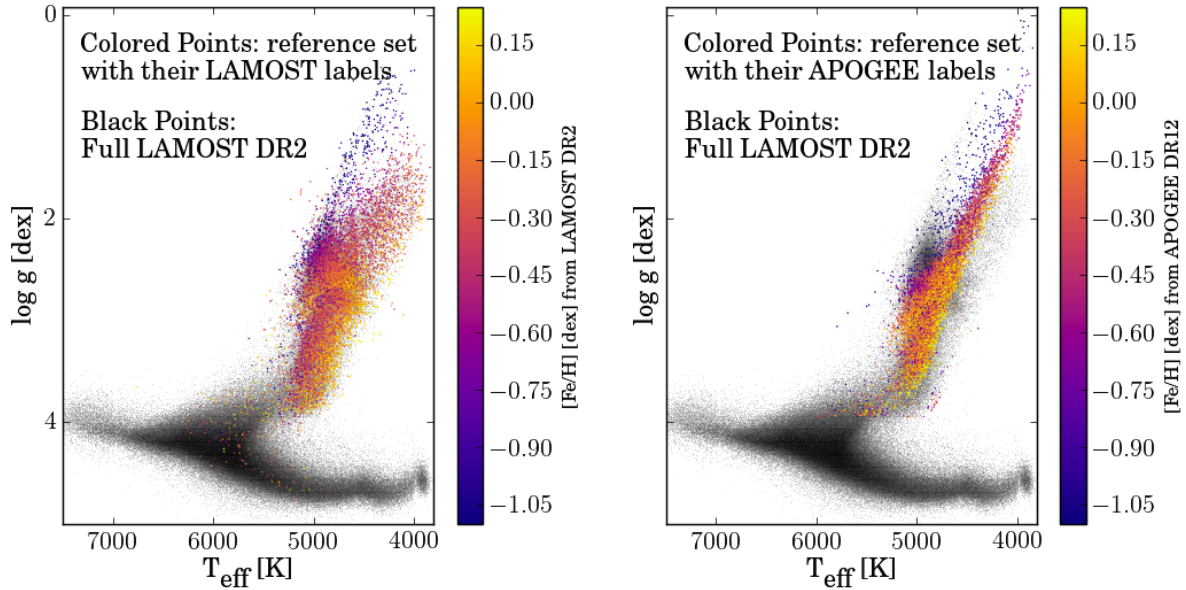

**Figure 3.** LAMOST DR2 (black points), overlaid with the reference set of 9952 objects (colored points) used to train the spectral model. These colored points are objects that have been observed by both LAMOST and APOGEE; in the left panel, they are shown with their LAMOST pipeline values, and in the right panel, they are shown with their APOGEE pipeline values. It is the values in the right panel that are used to train the spectral model.

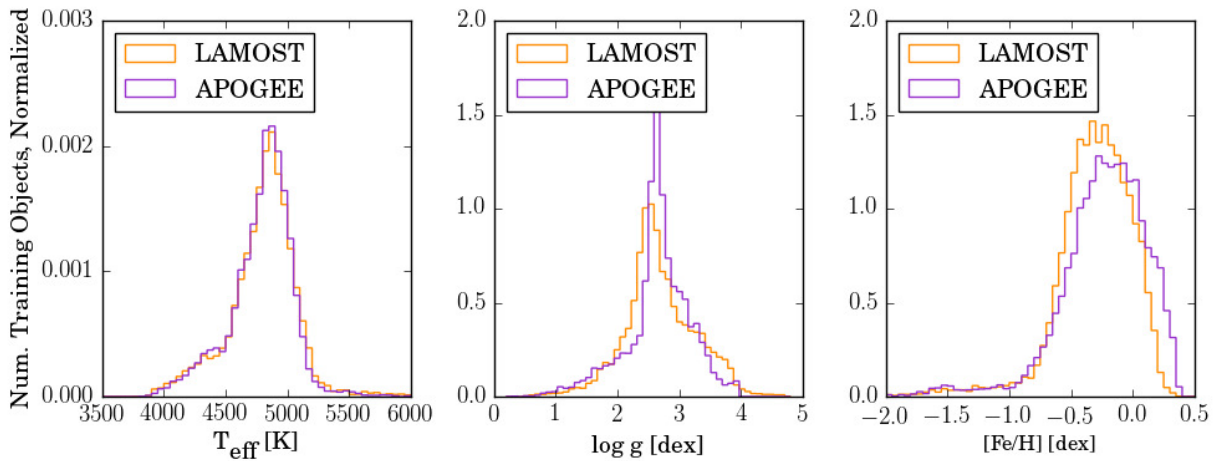

**Figure 4.** The distribution of labels for the 9952 training objects, values from LAMOST DR2 in yellow and values from APOGEE DR12 in purple. The purple (APOGEE) values are used to train the spectral model.

Following Ness et al. (2015) we presume that the model $g$ can be written as a linear function of $\ell_n$:

$$f_{n\lambda}^L = \boldsymbol{\theta}_\lambda^T \cdot \boldsymbol{\ell}_n^A + \text{noise} \tag{4}$$

corresponding to the single-pixel log likelihood function

$$\ln p(f_{n\lambda}^L \,|\, \boldsymbol{\theta}_\lambda^T, \boldsymbol{\ell}_n^A, s_\lambda^2) = -\frac{1}{2}\frac{[f_{n\lambda}^L - \boldsymbol{\theta}_\lambda^T \cdot \boldsymbol{\ell}_n^A]^2}{s_\lambda^2 + \sigma_{n\lambda}^2} - \frac{1}{2}\ln(s_\lambda^2 + \sigma_{n\lambda}^2) \quad . \tag{5}$$

For this work, once more as in Ness et al. (2015), we use a quadratic model such that $\boldsymbol{\ell}_n$ is

$$\begin{aligned}\boldsymbol{\ell}_n^A \equiv \Big[&1, \text{T}_\text{eff}, \log\text{g}, [\text{Fe/H}], [\alpha/\text{M}], \text{A}_\text{k}, \\
&\text{T}_\text{eff} \cdot \log\text{g}, \text{T}_\text{eff} \cdot [\text{Fe/H}], \text{T}_\text{eff} \cdot [\alpha/\text{M}], \text{T}_\text{eff} \cdot \text{A}_\text{k}, \log\text{g} \cdot [\text{Fe/H}], \log\text{g} \cdot [\alpha/\text{M}], \\
&\log\text{g} \cdot \text{A}_\text{k}, [\text{Fe/H}] \cdot [\alpha/\text{M}], [\text{Fe/H}] \cdot \text{A}_\text{k}, [\alpha/\text{M}] \cdot \text{A}_\text{k}, \\
&\text{T}_\text{eff}^2, \log\text{g}^2, [\text{Fe/H}]^2, [\alpha/\text{M}]^2, \text{A}_\text{k}^2\Big]_\text{SurveyX}\end{aligned} \tag{6}$$

The training step thus consists of holding the labels in the label vector $\boldsymbol{\ell}_n^A$ fixed (these are the reference labels) and optimizing the log likelihood to solve for the coefficients $[\theta_\lambda, s_\lambda^2]$ independently at every pixel. For a fixed scatter value, optimization is a pure linear-algebra operation (weighted least squares). Currently, we optimize for the scatter by stepping through a grid of scatter values.

Figure 5 shows the leading (linear) coefficient for each label as a function of wavelength, as well as the scatter as a function of wavelength. The magnitude of the leading coefficient can be thought of as the sensitivity of a particular pixel is to that particular label. Thus, Figure 5 is a way to visualize which regions of the spectrum are (as determined by *The Cannon*) important for which labels. We find that $\text{T}_\text{eff}$, log g, [Fe/H], and [$\alpha$/M] all have strong sensitivity to well-known spectral features such as Mg I, Na I D, and the Ca II triplet.

Interestingly, we find that $\text{A}_\text{k}$ has strong sensitivity not only to the Na I D doublet, but also to features that correspond to known diffuse interstellar bands (DIBs). The strongest of these DIBs are indicated by the orange lines in the lower panels of Figure 5. DIBs are absorption features that appear to arise from diffuse interstellar material; see Sarre (2006) and Herbig (1995) for extensive reviews. Over four hundred have been detected to date, mostly at optical wavelengths, but their origin remains uncertain (Hobbs et al. 2008; Herbig 1993). DIB strength has been found to correlate well with extinction and the column density of neutral hydrogen (Friedman et al. 2011). In addition, some DIBs seem to have correlated strengths, which suggests a shared origin (McCall et al. 2010; Friedman et al. 2011). Large-scale studies of DIBs (e.g. Yuan & Liu (2012)) hold promise for learning not only about their origin but also for mapping their environment; Zasowski et al. (2015) used DIBs in APOGEE infrared spectra to find that DIB strength is linearly correlated with extinction and thus a powerful probe of the structure and properties of the ISM. It is therefore perhaps not surprising that *The Cannon* learned to associate $A_k$ with DIB strength; features in the leading coefficients plot include well-known DIBs, e.g. at 4428 Å, 4882 Å, 5780 Å, 5797 Å, 6203 Å, 6283 Å, 6614 Å, and 8621 Å. Note that the DIBs in the Cannon model are effectively smeared across the radial velocity dispersion of the training sample.

## 4. *The Cannon* TEST STEP: DERIVING NEW STELLAR LABELS FROM LAMOST SPECTRA

In the training step (Section 3) we treated the labels from APOGEE $\ell_n^A$ as known and solved for the coefficients $\theta_\lambda$ of the spectral model. Now, in the test step, we take these spectral model coefficients and solve for new labels $\ell_n^L$ (as opposed to $\ell_n^A$) based on the spectra $f_{n\lambda}^L$ for each test object $n$. For a model that is quadratic in the labels, like ours, this consists of non-linear optimization. We use Python's `curve_fit` routine from the `scipy` library, which uses the Levenberg-Marquardt algorithm. We use seven starting points in label space, to assure convergence.

Before deriving new stellar labels for LAMOST objects, we test our model using a "leave-$\frac{1}{8}$-out" cross-validation test. We split the 9952 reference objects into eight groups, by assigning each one a random integer between 0 and 7. We leave out each group in turn, and train a model on the remaining seven groups. We then apply that model to infer new labels for the group that was left out. At the end of this process, each of the 9952 reference objects has a new set of labels determined by *The Cannon*, from a model that was *not* trained using that object.

### 4.1. *Cross-Validation*

Figure 6 shows the results of cross-validation. It shows four labels ($T_{\rm eff}$, log g, [Fe/H] and [$\alpha$/M]) determined by *The Cannon* directly from LAMOST spectra, plotted against the corresponding APOGEE (reference) labels, which were determined by ASPCAP directly from APOGEE spectra. For completeness, we show the output for extinction in the final panel (light purple). Note that, in this work, we consider extinction as a "nuisance" label: we fit for it in order to more reliably determine the four other labels, but the question of how to use *The Cannon* to reliably determine extinction values from spectra is beyond the scope of this work.

The low scatter and bias in the [$\alpha$/M] panel (bottom right) shows how well *The Cannon* transferred a *new* label to the LAMOST data set. The scatter in all four labels for the objects with S/N > 50 LAMOST spectra (roughly half of the objects) is comparable to the typical uncertainties from ASPCAP, which are 91.5 K in $T_{\rm eff}$, 0.11 in log g, and around 0.05 in both [Fe/H] and [$\alpha$/M] (Holtzman et al. 2015). (To clarify, the model was trained on and applied to objects of all S/N; we are simply quoting scatter values for objects with S/N > 50. The dependence of scatter with S/N is shown in Figure 8.) Note that the scatter in [$\alpha$/M] derived from the LAMOST spectra is very similar to the precision in [α/Fe] inferred indirectly for the SEGUE G-dwarfs by Bovy et al. (2012), based on SDSS spectra at similar resolution, wavelength coverage and S/N. Note also that the discontinuity in [α/M] is present in the reference set (because of the existence of two physical alpha sequences, the alpha-enhanced and alpha-poor sequences) and recovered in the test step, despite the fact that the model itself is in no way bimodal. The model is a quadratic function: nothing about it encourages a separation of these populations. Thus, this represents further physical verification of the model's accuracy.

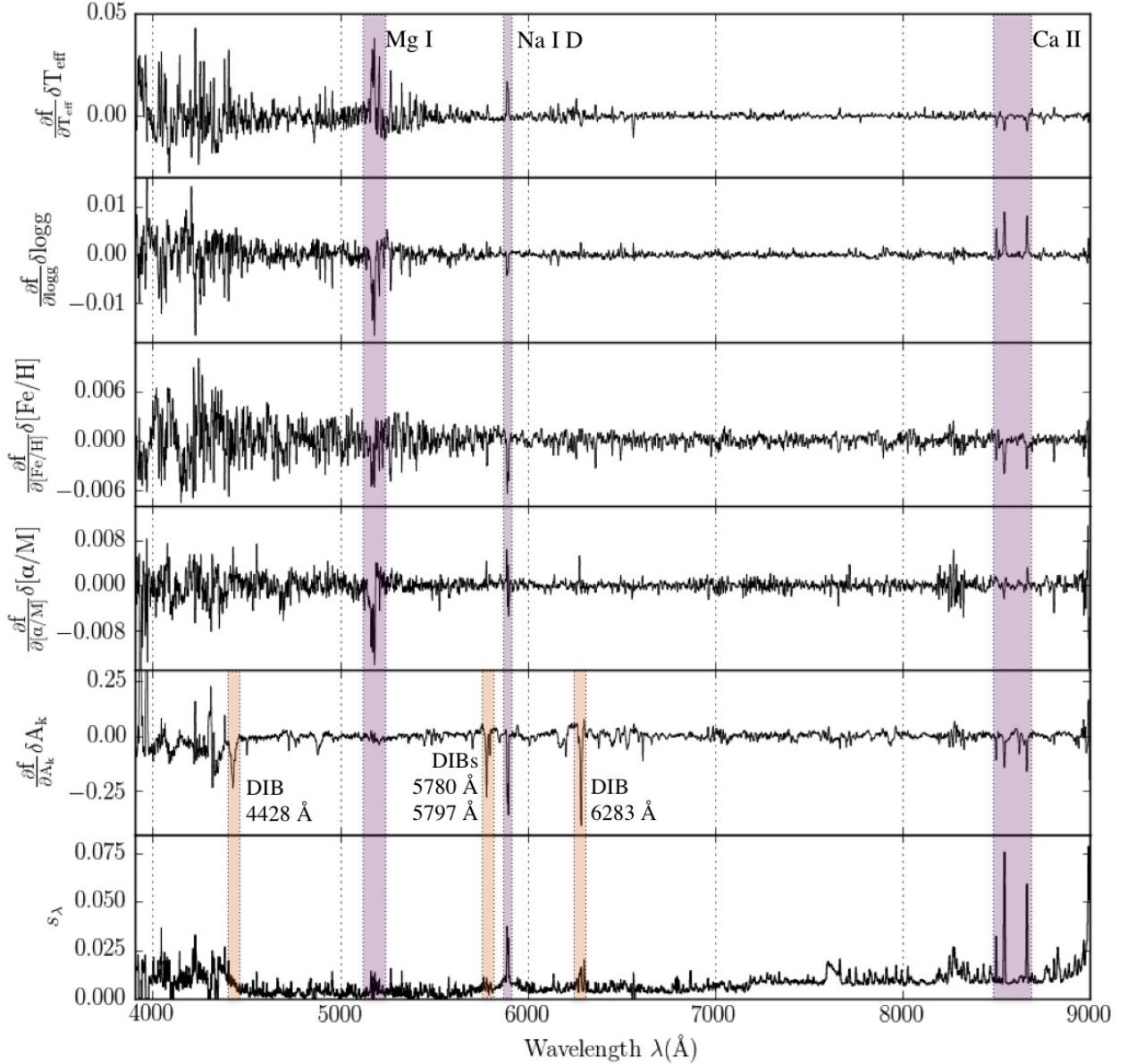

**Figure 5.** Leading (linear) coefficients and scatter from the best-fit spectral model, with prominent features labeled. These coefficients indicate how sensitive each pixel in the spectrum is to each of the labels. In the top four panels, note peaks at well-known spectral features such as the Mg I triplet around 5170 Å and the Ca II triplet around 8600 Å. In the fifth panel, note peaks at well-known diffuse interstellar bands (DIBs). The coefficients are scaled by the approximate errors in the labels (91.5 K in $T_{\mathrm{eff}}$, 0.11 in $\log g$, 0.05 in [Fe/H] and [α/M]; Holtzman et al. (2015)).

This information is represented as residuals in Figure 7; a direct comparison with Figure 1 shows a significant improvement in scatter and a dramatic reduction of systematic differences between the labels derived from LAMOST and APOGEE spectra, particularly in $\log g$ and [Fe/H]. The inter-survey biases in the three labels have all but vanished, demonstrating that we have successfully measured APOGEE-scale labels directly from LAMOST spectra, thus bringing the two surveys onto the same scale. Note also, that the scatter (at a given S/N) has been reduced considerably: *The Cannon* can also measure more precise labels from the low-resolution LAMOST spectra (Ness et al. 2015).

In both Figure 6 and Figure 7, there is a clear turn-off at low temperatures, $T_{eff} \lesssim 4250$. Our model in this regime is limited by the fact that ASPCAP labels are less reliable at these lower temperatures, so we urge caution when using labels for objects at lower temperatures. We return to this in Section 4.2 and Section 5.

Furthermore, *The Cannon* performs more precisely at low S/N than the LAMOST pipeline, as seen in Figure 8. Here, for a S/N metric, we define "$\sim SNR_g$." We quantify S/N in the g-band because the leading coefficients show that decisive information comes from this regime. Furthermore, the error bar and S/N should reflect the variance of each pixel around the best-fit model; thus, the $\chi^2$ of a model that fits well (in this case, the model from *The Cannon*) should roughly equal the number of pixels in the spectrum, 3626. Instead, the $\chi^2$ led us to find that the errors and S/N in the spectra needed to be adjusted by a factor of three. Thus, $\sim SNR_g$ represents the S/N in the g-band, multiplied by three.

Figure 9 provides verification that the label transfer in $T_{eff}$ and $\log g$ has led to astrophysically plausible results. It compares the ($T_{eff}$, $\log g$) distribution for all reference objects using their labels from the APOGEE pipeline, from the LAMOST pipeline, and from the Cannon model for the LAMOST data. Both the morphology of the red clump and of the giant branch shows that the Cannon labels are physically much more plausible than the pipeline labels derived from the same LAMOST data.

Finally, the "goodness of fit" can be quantified by a $\chi^2$ value that takes into account uncertainty in the data and scatter in the model. This $\chi^2$ essentially amounts to a comparison between the model spectrum and the data. This is visualized in Figure 10, which compares the data to the Cannon model spectrum for a randomly selected LAMOST object, centered on the Mg I triplet. The spectra line up nearly perfectly, to within the uncertainties in the data and scatter in the model. This demonstrates that the model, with the five labels we are fitting for, is an excellent description of LAMOST spectra. The success of cross-validation motivates and justifies the application of the model to LAMOST objects that have *not* been observed by APOGEE.

### 4.2. *Application to LAMOST DR2*

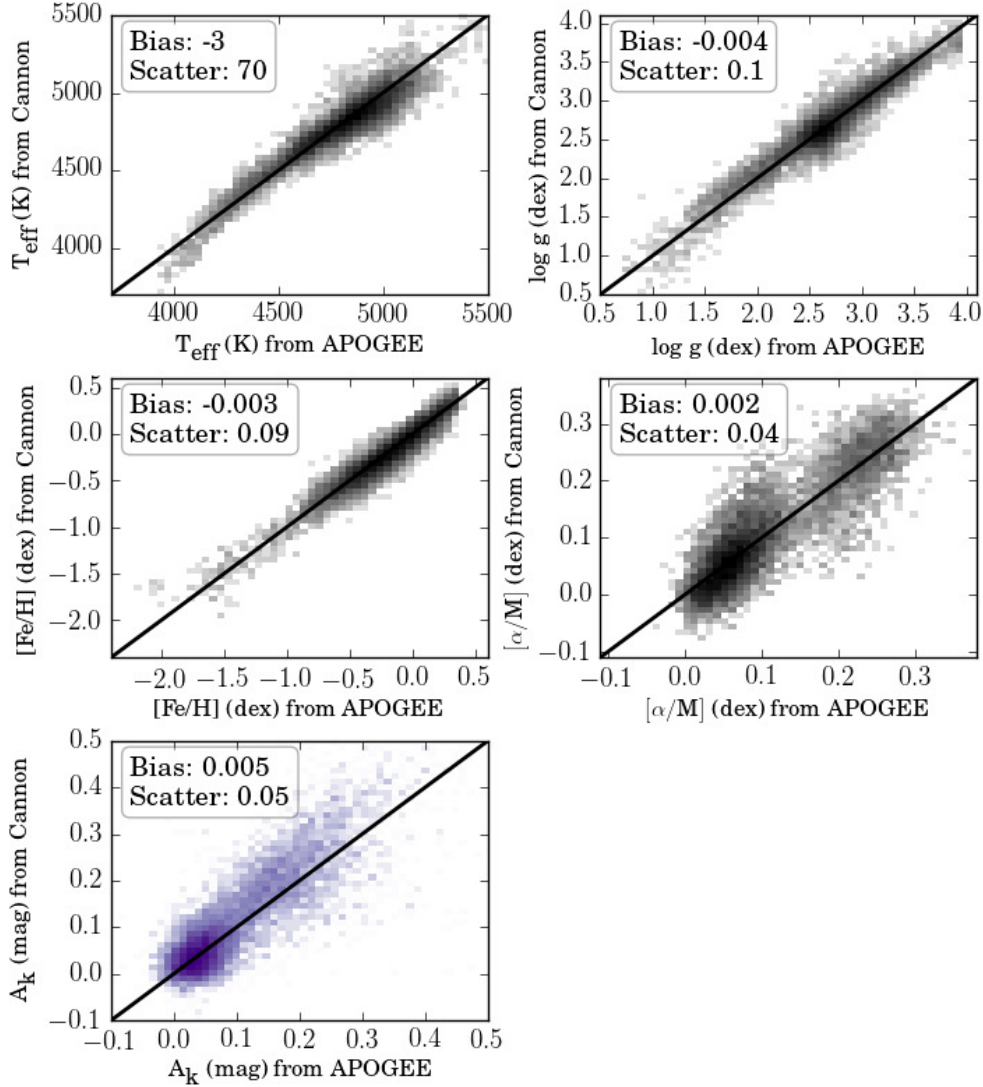

**Figure 6.** Cross-validation of *The Cannon*'s label transfer from APOGEE to LAMOST : Shown are the APOGEE labels of all reference objects compared to the labels derived from LAMOST data by *The Cannon* in the test step. We emphasize that no object in this figure was used to train the model that inferred its labels. The tight one-to-one correlations in the $T_{eff}$, $\log g$ and [Fe/H] panels reflect the quality of the label transfer. The bottom right panel shows how well *The Cannon* is able to transfer the *new* label [$\alpha$/M] from APOGEE. The success with which cross-validation reproduces the reference labels serves to justify our application of this method to a more extensive LAMOST sample. For completeness, we include extinction as a fifth panel, but emphasize that ours is not a reliable method for inferring extinction from LAMOST spectra. The scatter and bias values represent spectra with S/N> 50.

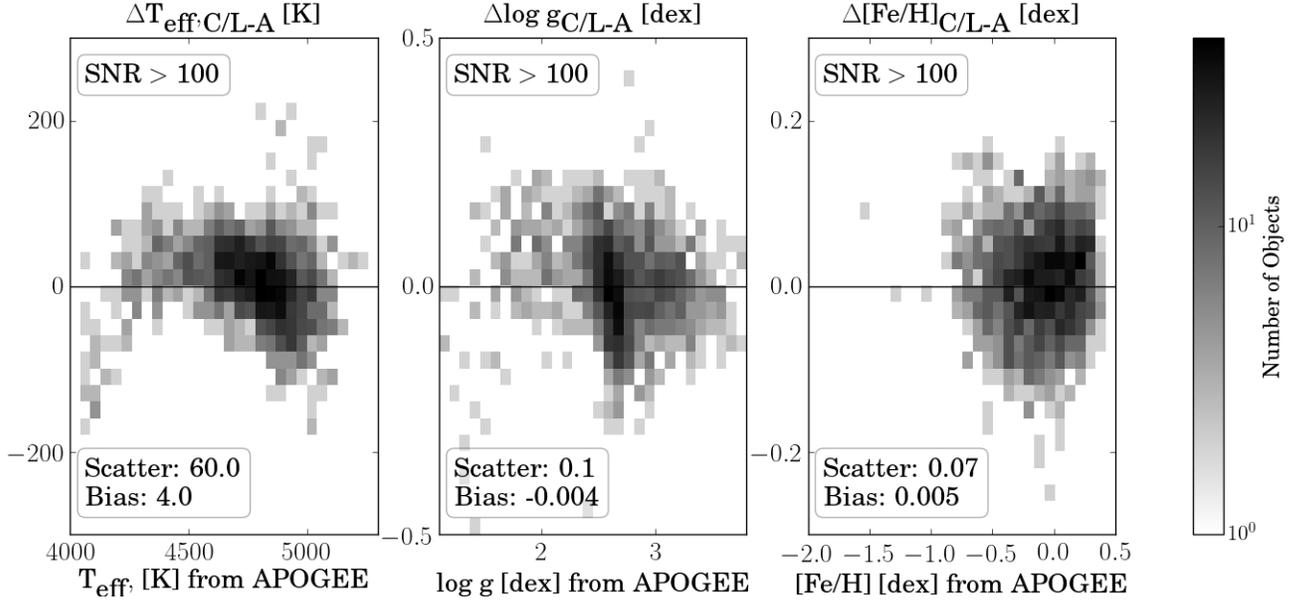

**Figure 7.** Comparison between *The Cannon* output and APOGEE reference labels : Shown here are labels for the 9952 in the reference set, objects measured in common between LAMOST and APOGEE. The systematic differences between labels determined by *The Cannon* from LAMOST spectra and by ASPCAP from APOGEE spectra have been almost completely eliminated (see (Figure 1). The Cannon values also show a substantially reduced scatter with respect to the APOGEE-labels, presumed to be ground-truth here.

We now turn to applying the spectral model to DR2 objects that were *not* observed by APOGEE. *The Cannon* cannot extrapolate to regimes of ($T_{eff}$, log g, [Fe/H], [$\alpha$/M]) label space that are completely different from those represented in the reference set, as shown in Ness et al. (2015). We believe that it is the *bounds* of the training labels that limit the applicability of the model, rather than the *distribution* of the training labels. This is because the label distribution is not sparse; the reference set densely populates the training label space (see Figures 3 and 4). In addition, the model is quadratic and is therefore fit smoothly across the label space.

So, we restrict our test set to LAMOST DR2 objects that are reasonably close to the reference set in label space. To do so, we define a "label-distance" $D$ from the reference objects in label space, exploiting here that all test objects have (initial) stellar label estimates from the LAMOST pipeline. The label-distance of a LAMOST test object (in LAMOST label space; subscript $L$) and a reference object (in APOGEE label space; subscript $A$) is

$$D = \frac{1}{K_{T_{eff}}^2}(T_{eff,L} - T_{eff,A})^2 + \frac{1}{K_{\log g}^2}(\log g_L - \log g_A)^2 + \frac{1}{K_{[Fe/H]}^2}([Fe/H]_L - [Fe/H]_A)^2, \quad (7)$$

where we have normalized by the approximate uncertainty in each label: $K_{T_{eff}} = 100$, $K_{\log g} = 0.20$, and $K_{[Fe/H]} = 0.10$. We then calculate an object's label-distance from the reference set by taking the average of its label-distances to the ten nearest reference objects.

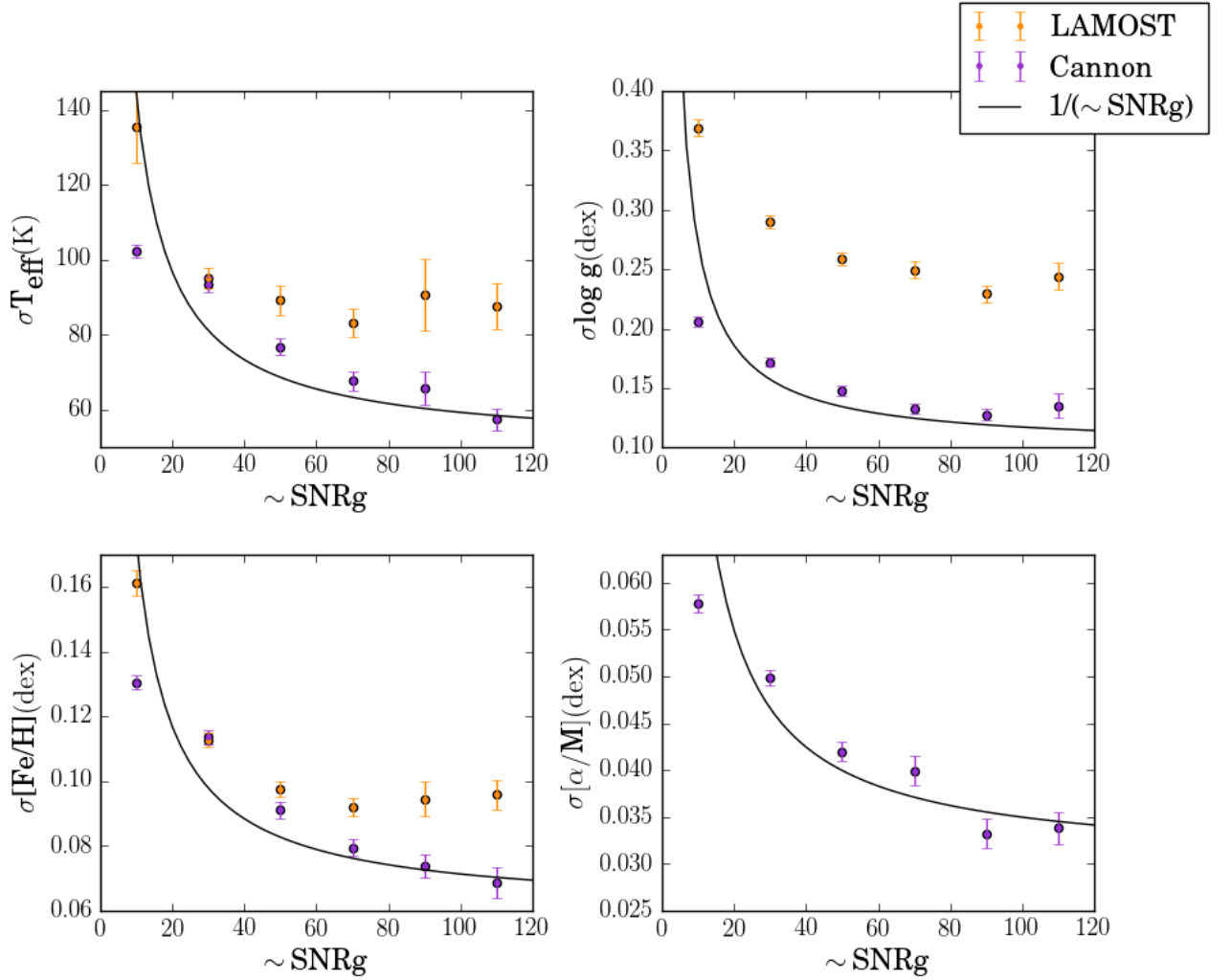

**Figure 8.** The S/N-dependence of the scatter between APOGEE DR12 labels and the corresponding labels measured from LAMOST spectra by *The Cannon* (purple points) and LASP (yellow points). *The Cannon* represents a substantial improvement from the LAMOST pipeline in the three labels that the the APOGEE and LAMOST pipelines measure in common, and the model behaves well with decreasing S/N. The performance improvement is generally steeper than the inverse of the S/N. Note that we are using our own value for $\sim \mathrm{SNR_g}$, which does not reflect the reported LAMOST error bar.

We use these label-distances to define the regime within which a LAMOST DR2 object was deemed a feasible test object. The label-distance cut was determined by running the test step of *The Cannon* on 3,000 random objects in LAMOST DR2. This showed that there is a particular label-distance (roughly 2.5, as defined in Equation 7) at which the giant branch and main sequence populations separate: since the reference set comprises only giants, stars on the giant branch are closer to the reference set in label space than stars on the main sequence. As expected, running these 3,000 objects through the test step (using *The Cannon* to try and reproduce their reference labels) demonstrated that *The Cannon* was better able to reproduce the training labels for stars within this label-distance than for stars outside this label-distance.

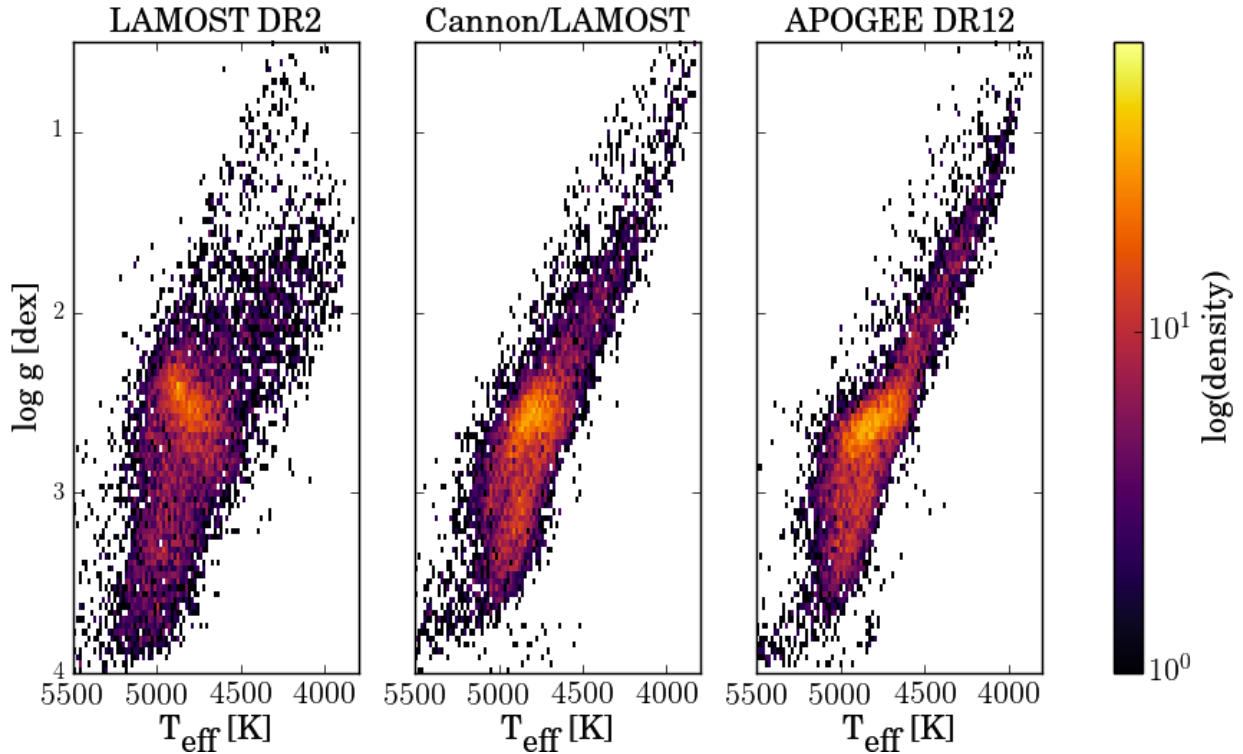

**Figure 9.** Astrophysical verification of the labels derived by *The Cannon* model for LAMOST data: the panel show the distribution of all reference objects in the ($T_{eff}$, $\log g$) plane, using their LAMOST DR2 labels (left), Cannon labels from LAMOST spectra (center), and APOGEE DR12 labels (right). The distribution of Cannon labels is not only much more similar to ASPCAP's labels, but also much more physically plausible, exhibiting a tighter red clump and a more well-defined upper giant branch.

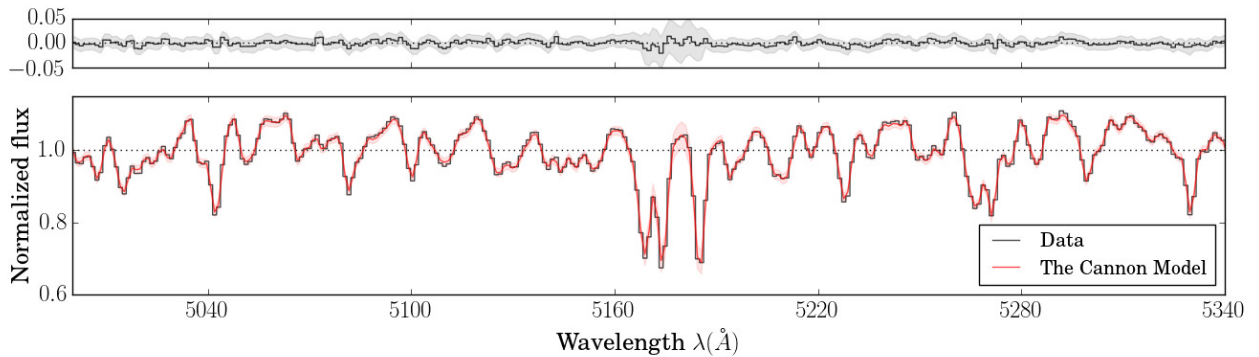

**Figure 10.** A sample model spectrum: a portion of the (Cannon-)normalized spectrum for a randomly selected star in the validation set, centered on the Mg I triplet. The best-fit model spectrum is in red and the data is in black. The residuals are plotted in the top panel. To emphasize, this object was *not* used to train the model that inferred its labels.

Thus, we use this label-distance cut to inform our choice of test objects: we select those with a label distance to the reference set of less than 2.5. Effectively, this is a way to select only giants; we are restricted to giants because these happen to be the objects with reference labels. Figure 11 shows 14,000 random stars in the ($T_{\rm eff}$, log g)-plane (colored points), on top of the entire LAMOST DR2 sample (see Figure 3): a label-distance cut at 2.5 neatly separates the giants (to which the spectral model applies) from the main sequence stars.

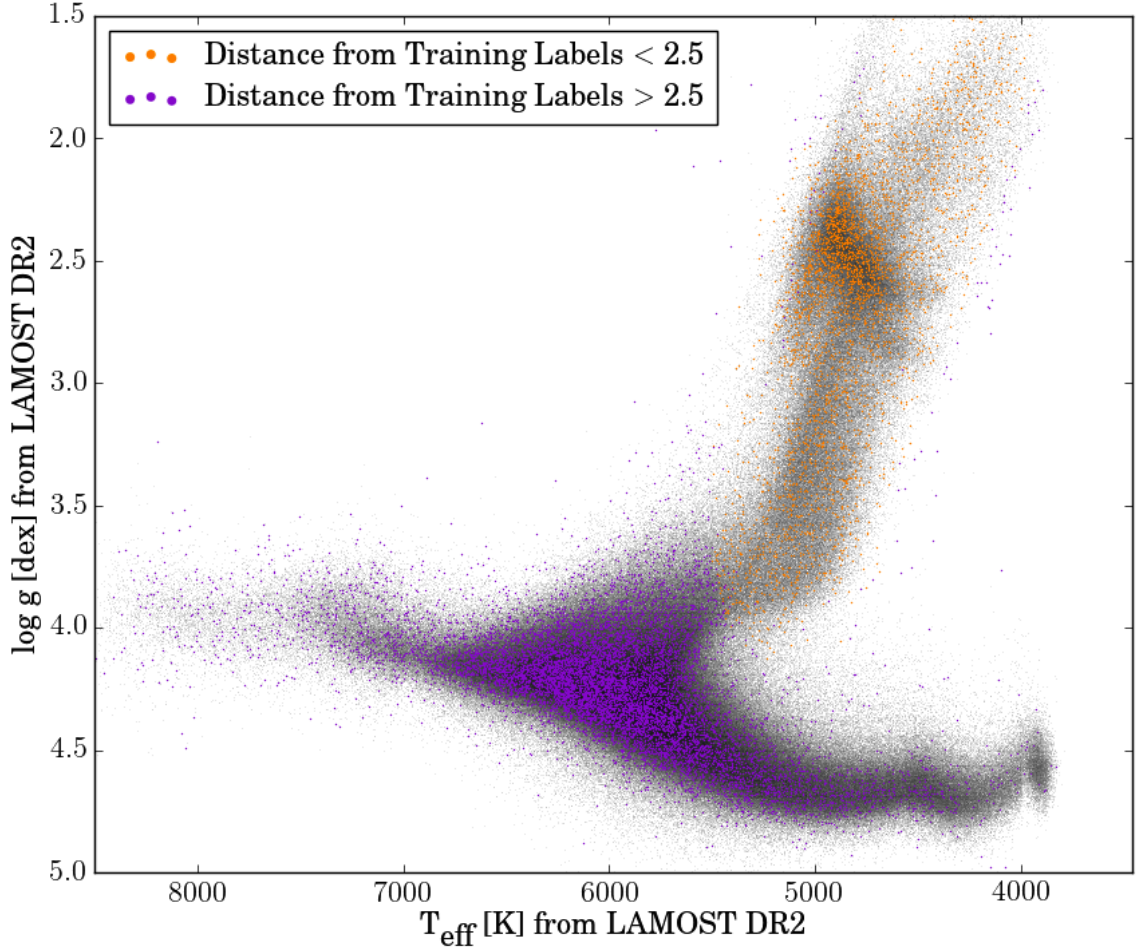

**Figure 11.** Label Distance From Reference Set: (Black) all LAMOST DR2 points in ($T_{\rm eff}$,log g) space, with (Color) 14,000 objects overlaid color-coded by their distances from the reference label space. For the test step, we choose objects whose distances are less than 2.5, which amounts to effectively selecting giant stars. The fact that our reference set consists only of giants restricts the applicability of our model to this regime; that is, in this work we only apply the model to the orange points. The structure seen in the lower main sequence is presumably an artifact of the LAMOST analysis.

We define the test set as all LAMOST DR2 objects with a label-distance from the reference set of $< 2.5$. After using the spectral model to infer new labels, we excise objects for which the convergence either failed or resulted in a fit with reduced $\chi^2 > 10$ (fewer than 0.1% of the objects). This leaves 444,228 stars (giants), not including the reference set. Figure 12 shows the $T_{\rm eff}$, log g plane for a 44,000 of these objects (those within the window -0.1 $<$[Fe/H] $<$0.0, chosen simply for clarity). The values measured by *The Cannon* form a much cleaner and more well-defined red clump and upper giant branch, illustrating the improvement in precision we have achieved by transferring labels from APOGEE.

The newly inferred labels for all of 444,228 objects, together with the *Cannon* labels for the reference set and their associated formal uncertainties (from the covariance matrix), are available in a table online and an excerpt is shown in Table 1. This table includes the first $[\alpha/M]$ values measured for the full set of LAMOST giants. We note that $[\alpha/M]$ has been previously measured for roughly 1% of this sample, using the SEGUE pipeline (Lee et al. 2015). As mentioned previously, this could be done because SEGUE and LAMOST have very similar resolution and wavelength coverage. We emphasize that the approach with *The Cannon* does not rely on surveys having any overlapping wavelength coverage, or even comparable spectral resolution.

We further emphasize that at lower temperatures, $T_{\rm eff} \lesssim 4250$, our model is less reliable, as shown in Figures 6 and 7. Thus, we urge caution when using the catalog for objects in this temperature regime, which is roughly 3% of the sample.

In addition to the formal uncertainties from the covariance matrix, there are a number of sources of uncertainty that we now address. First, the discreteness (that is, the incomplete and sparse coverage) of the reference set induces an uncertainty in the final estimation of the labels. To estimate the strength of this effect, we create 20 different spectral models by bootstrap-sampling from the set of reference objects. For each set, we run the cross-validation as described in Figure 6. A subset of the test set has 20 different label estimates, and we adopt the standard deviation of these measurements to reflect the uncertainties of these new LAMOST labels. With such a large training sample with which to fit the spectral model, the values are negligible: 4.4 K for $T_{\rm eff}$, 0.012 dex for $\log g$, 0.0060 dex for [Fe/H], and 0.0042 dex for $[\alpha/M]$.

Furthermore, there is a contribution from the uncertainties in the labels used to train the model, which we do not account for in this version of *The Cannon*. Thus, although we only report the formal uncertainty in Table 1, the number is certainly an underestimate. The spread in the cross validation (see Section 4.1, Figure 6, and Figure 8) provides an estimate of the uncertainties. It is important to recall, however, that "uncertainty" here is the departure from the APOGEE value. Our goal is to make measurements consistent with the APOGEE scale; we cannot improve upon the accuracy of the reference system.

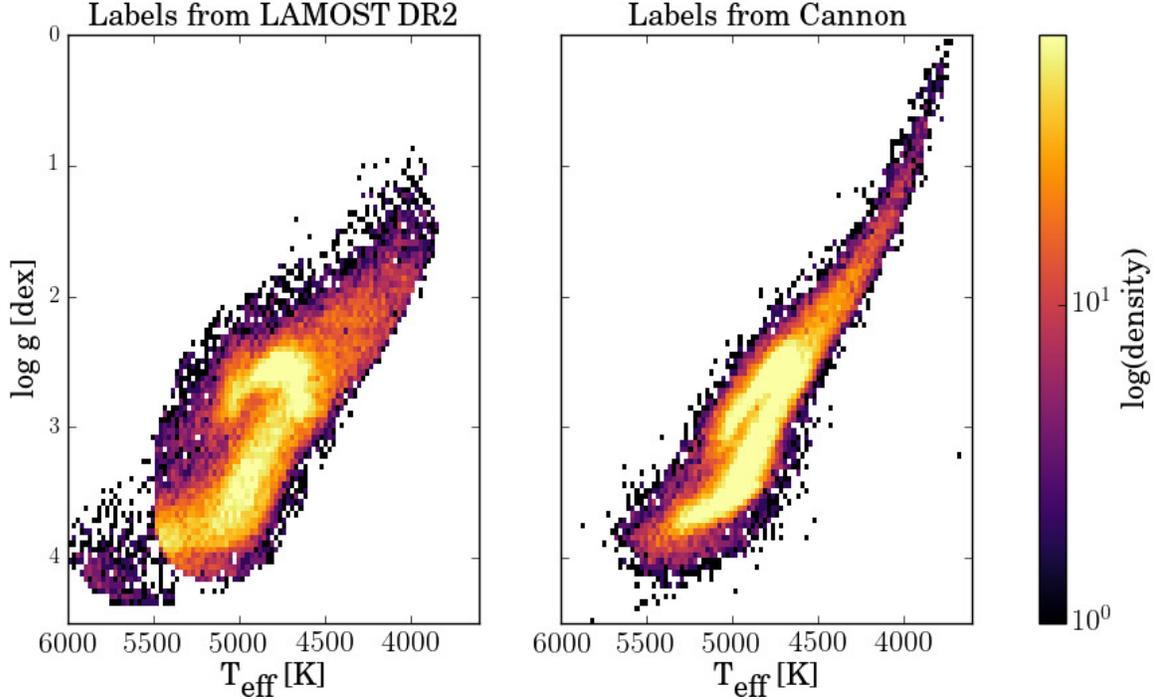

**Figure 12.** Precision of the new labels: the $T_{eff}$, $\log g$ plane for the $\sim$44,000 test objects in a narrow [Fe/H] window: $-0.1 <$ [Fe/H] $< 0.0$. The left panel shows the values from the LAMOST DR2 catalog, as determined by LASP. The right panel shows the values from *The Cannon* applied to the same spectra. The Cannon labels are clearly much more precise and astrophysically plausible, exhibiting a more well-defined red clump and upper giant branch.

**Table 1.** Excerpt from the online table of stellar labels ($T_{eff}$, $\log g$, [Fe/H], and [$\alpha$/M]) for 454,180 stars, inferred by *The Cannon* directly from LAMOST spectra. Column 1 is the unique LAMOST FITS Name of the object in the format spec$-$<lmjd>$_-$<spid>$-$<fiberid>, which can be resolved at the LAMOST DR2 website[2]. Columns 2-5 are the labels from *The Cannon*, Columns 6-9 are the formal errors on those labels from the covariance matrix in the Cannon model fit, and the final column is the reduced $\chi^2$. Note that the reduced $\chi^2$ values are low by a factor of $\sim 3$ because the random component of the errors in the LAMOST spectra is overestimated (see Section 4).

| LAMOST ID | $T_{eff}$ (K) | $\log g$ (dex) | [Fe/H] (dex) | [$\alpha$/M] (dex) | $\sigma(T_{eff})$ (K) | $\sigma(\log g)$ (dex) | $\sigma$([Fe/H]) (dex) | $\sigma$([$\alpha$/M]) (dex) | Red. $\chi^2$ |
|---|---|---|---|---|---|---|---|---|---|
| spec-55859-F5902_sp01-034 | 4899 | 3.15 | -0.597 | 0.207 | 48 | 0.08 | 0.053 | 0.024 | 0.62 |
| spec-55859-F5902_sp01-136 | 5279 | 3.08 | -0.838 | 0.206 | 145 | 0.29 | 0.177 | 0.075 | 0.57 |
| spec-55859-F5902_sp01-202 | 4884 | 3.25 | -0.383 | 0.225 | 36 | 0.06 | 0.040 | 0.016 | 0.82 |
| spec-55859-F5902_sp01-207 | 4882 | 3.43 | -0.252 | 0.186 | 39 | 0.06 | 0.043 | 0.017 | 0.88 |

### 4.3. *The [$\alpha$/M] Map of the Milky Way from LAMOST*

The full astrophysical verification and exploitation of the new set of labels for the LAMOST DR2 giants is beyond the scope of the paper. Here, we give some initial indication of what will be enabled, by showing the ([Fe/H],[$\alpha$/M]) plane (Figure 13) and the distribution of [$\alpha$/M] in galactic latitude and longitude (Figure 14) for all LAMOST DR2 giants. This is by far the largest set of giants with the $[\alpha/M]$ abundance label. As Fig. 14 shows, the combination of the two surveys overcomes a limitation of many previous analyses of the abundance-dependent Galactic disk structure (see e.g. Rix & Bovy (2013)): most large surveys have either extensive coverage at high Galactic latitudes with sparse sampling in the Galactic plane, or vice versa. The distribution in the ([Fe/H],$[\alpha/M]$) plane looks very plausible, exhibiting the $\alpha$-enhanced and the low-$\alpha$ sequences, and the spatial distribution beautifully exhibits the low-alpha, chemically late, young population in the mid-plane and at large radii, and the alpha-enhanced, rapidly enriched, old population in the thick disk (high latitudes) and Galactic center.

As this represents the first (and only) attempt to measure $[\alpha/M]$ for most of these objects, we cannot prove that these values are "correct" in an absolute sense. In particular, we cannot know whether the test set falls within the $[\alpha/M]$ range of the reference set, or whether *The Cannon* is extrapolating outside the $[\alpha/M]$ range of the reference set. We do not believe that this is a significant issue, as spectra should be dominated by $T_{\text{eff}}$, log g, and [Fe/H]. This is supported by Figures 13 and 14 and the fact that, in the test step, the model does an excellent job of predicting the spectra as quantified by the low $\chi^2$ values and visualized in Figure 10. Our paper stands as the only prediction of this label - and the best one that can be made by *The Cannon* given the available overlap (training) set. We encourage future observations to test these predictions.

In addition, *The Cannon* is certainly not the only possible way to measure alpha-enhancement values from LAMOST spectra, and this may not be the best measurement possible with *The Cannon*. In particular, allowing the model to fit for extinction via DIBs in the spectrum may be problematic, because the DIBs are in a different velocity frame from the star. In a future paper, it may be worth exploring masking the DIBs from the spectrum. For now, we simply seek to demonstrate that alpha-enhancement values *can* be measured from these spectra using a data-driven technique to transfer values from the APOGEE label system. Unlike traditional cross calibration methods, a method like *The Cannon* that transfers information from one survey to another does not rely on both survey pipelines measuring a set of parameters; we can measure alpha enhancement despite the fact that the LAMOST pipeline has not attempted to measure those values, because we build a spectral model directly from APOGEE data.

## 5. DISCUSSION

We have demonstrated that *The Cannon* can be used to put two spectroscopic stellar surveys with very different experimental set-ups (wavelength coverage, resolution) onto the same label (stellar parameter and chemical abundance) scale by training a spectral model on the set of objects observed in common between the surveys. We used LAMOST and APOGEE as our example, and showed that we can greatly reduce the systematic differences between labels measured using the two individual survey pipelines. We can also boost the precision of the label estimates for the data set of less resolution and S/N (LAMOST in this case). By training our model to infer APOGEE-scale stellar labels directly from LAMOST spectra, we can also transfer new labels from one survey to another for: here we derived [α/M] for the full set of LAMOST giants for the first time.

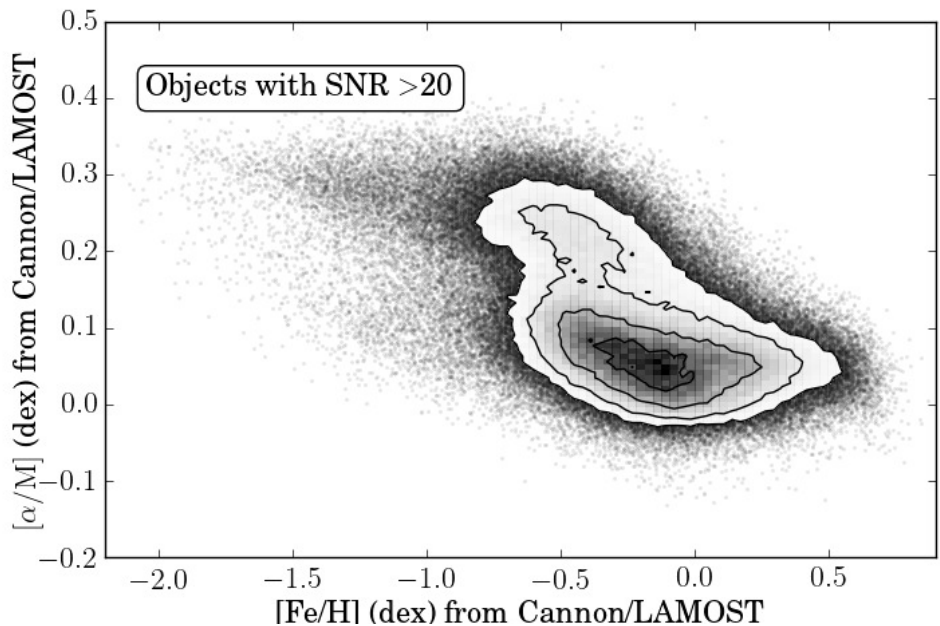

**Figure 13.** The ([Fe/H], [α/M]) plane, showing the labels determined by *The Cannon* for 305,694 of the 454,180 objects: those with LAMOST spectra S/N > 20. The raw values are shown as grayscale points and the contours (made from logarithmic bins) are at 0.5, 1, 1.5, and 2 sigma. These are the first [α/M] values measured for the full set of LAMOST giants, and by far the largest set of giants with this abundance label. Figure made using code from Dan Foreman-Mackey et al. (2014).

There are substantial benefits to using *The Cannon* for label transfer. As described in (Ness et al. 2015), *The Cannon* is very fast: for 9952 objects, on a regular computer, the training step took a few minutes and the test step for 444,228 test objects (i.e. the label determination) took a few hours. In addition, *The Cannon* requires no physical models and performs well at low S/N and low resolution: in this case, we were able to measure labels of comparable precision to APOGEE (at least, to ASPCAP's stated precision; see Ness et al. (2015)) from LAMOST's substantially lower resolution and lower S/N spectra (see Figure 8). Finally, because *The Cannon* fits for a set of model coefficients independently at each wavelength of the spectrum, there is a straightforward way to investigate the information content of a particular wavelength regime and determine where and how information about a particular label is encoded (see Figure 5).

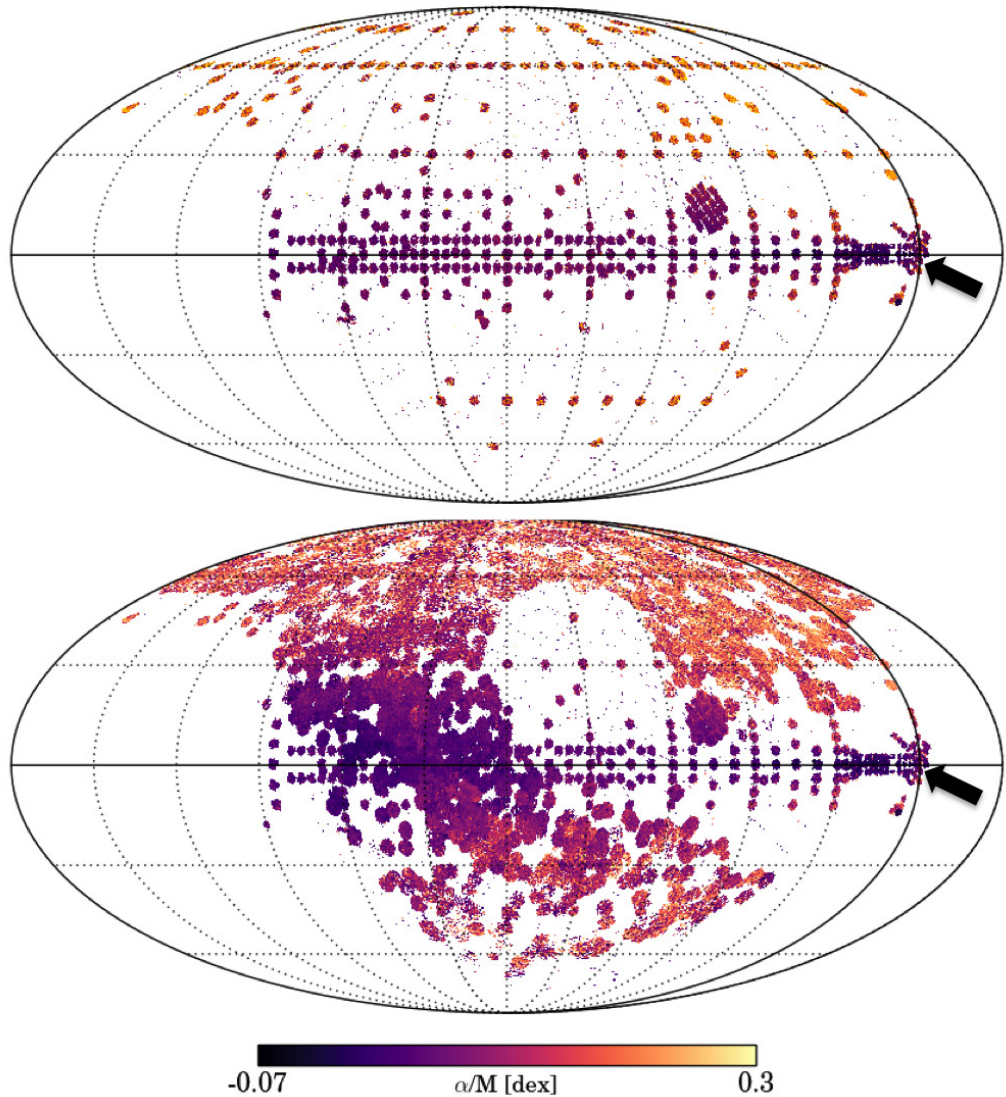

**Figure 14.** The distribution on the sky (in Galactic coordinates) of the full set of objects with consistently measured [$\alpha$/M]: the top panel shows the full APOGEE sample with $\approx$ 100,000 objects, and the bottom panel shows these values combined with 454,180 [$\alpha$/M] inferred by *The Cannon* from the LAMOST spectra. The much more extensive area coverage of the LAMOST data is immediately apparent. One can clearly see how the low-$\alpha$ stars, presumably a younger population from more slowly enriched gas, is concentrated towards the mid-plane. The $\alpha$-enhanced stars, mostly a rapidly enriched, old population, are found in the thick disk and halo (at high latitudes) as well as in the outer Galactic bulge; the arrow on the right denotes the Galactic center. This illustrates the promise of survey label transfer for stitching together a more complete stellar population picture of the Galaxy.

This label transfer effort was both enabled and severely limited by the reference set. The large number of objects with reliable labels (9952) measured in common between the two surveys enabled us to fit for a spectral model, but the incomplete label coverage restricted the applicability of the model to only 454,180, roughly 20% of the several million LAMOST objects. Furthermore, the quality of the reference labels at low $T_{\text{eff}}$ restricted our ability to reliably model spectra in that regime (see Section 4.1). To take full advantage a data-driven approach like *The Cannon*, it is essential for surveys to measure objects in common that have high-fidelity labels comprehensively spanning the label space of interest.

Clearly, *The Cannon* holds promise for bringing other overlapping surveys onto the same label scale (e.g. RAVE, SEGUE, GALAH, Gaia-ESO). Looking ahead, *Gaia* will provide a billion low-resolution spectra. By the time these spectra become available, over a million of these objects will have spectroscopic labels determined by much higher resolution ground-based spectra. This offers a tremendous opportunity for transferring high-quality spectral labels to low-resolution *Gaia* spectra, if not with the present version of *The Cannon* then with the basic underlying ideas of data-driven spectral modeling.

**Note added in revision:** after the completion and submission of our paper, Li et al. (2016) also demonstrated that [$\alpha$/M] can be realiably measured from LAMOST spectra. They developed a technique for this measurement using template matching and an extension of the LAMOST Stellar Pipeline.

The code used to produce the results described in this paper was written in Python and is available online in an open-source repository.[3] An archival copy has been preserved with Zenodo (Ho et al. 2016).

It is a pleasure to thank Jo Bovy (U. Toronto), Andy Casey (IoA Cambridge), Morgan Fouesneau (MPIA), Evan Kirby (Caltech), Branimir Sesar (MPIA), and Yuan-Sen Ting (Harvard) for valuable discussions and assistance. AYQH is grateful to the community at the MPIA for their support and hospitality during the period in which most of this work was performed. The authors would like to thank two anonymous referees for their detailed and constructive feedback, which greatly improved the strength and clarity of the paper.

AYQH was supported by a Fulbright grant through the German-American Fulbright Commission and a National Science Foundation Graduate Research Fellowship under Grant No. DGE1144469. MKN and HWR have received funding for this research from the European Research Council under the European Union's Seventh Framework Programme (FP 7) ERC Grant Agreement n. [321035]. DWH was partially supported by the NSF (grant IIS-1124794), NASA (grant NNX08AJ48G), and the Moore-Sloan Data Science Environment at NYU. CL acknowledges the Strategic Priority Research

---

[3] www.github.com/annayqho/TheCannon

Program "The Emergence of Cosmological Structures" of the Chinese Academy of Sciences, Grant No. XDB09000000, the National Key Basic Research Program of China 2014CB845700, and the National Natural Science Foundation of China (NSFC) grants No. 11373032 and 11333003.

Guoshoujing Telescope (the Large Sky Area Multi-Object Fiber Spectroscopic Telescope LAMOST) is a National Major Scientific Project built by the Chinese Academy of Sciences. Funding for the project has been provided by the National Development and Reform Commission. LAMOST is operated and managed by the National Astronomical Observatories, Chinese Academy of Sciences.

Funding for the Sloan Digital Sky Survey IV has been provided by the Alfred P. Sloan Foundation, the U.S. Department of Energy Office of Science, and the Participating Institutions. SDSS-IV acknowledges support and resources from the Center for High-Performance Computing at the University of Utah. The SDSS web site is www.sdss.org.

SDSS-IV is managed by the Astrophysical Research Consortium for the Participating Institutions of the SDSS Collaboration including the Brazilian Participation Group, the Carnegie Institution for Science, Carnegie Mellon University, the Chilean Participation Group, the French Participation Group, Harvard-Smithsonian Center for Astrophysics, Instituto de Astrofísica de Canarias, The Johns Hopkins University, Kavli Institute for the Physics and Mathematics of the Universe (IPMU) / University of Tokyo, Lawrence Berkeley National Laboratory, Leibniz Institut für Astrophysik Potsdam (AIP), Max-Planck-Institut für Astronomie (MPIA Heidelberg), Max-Planck-Institut für Astrophysik (MPA Garching), Max-Planck-Institut für Extraterrestrische Physik (MPE), National Astronomical Observatory of China, New Mexico State University, New York University, University of Notre Dame, Observatário Nacional / MCTI, The Ohio State University, Pennsylvania State University, Shanghai Astronomical Observatory, United Kingdom Participation Group, Universidad Nacional Autónoma de México, University of Arizona, University of Colorado Boulder, University of Oxford, University of Portsmouth, University of Utah, University of Virginia, University of Washington, University of Wisconsin, Vanderbilt University, and Yale University.

*Facilities:*

*Facility:* Sloan(APOGEE spectrograph),

*Facility:* LAMOST